\documentclass[apj,iop]{emulateapj}
\usepackage{natbib}
\usepackage{graphicx}

\usepackage{tikz}
\usepackage{bm}
\usepackage{amsmath}
\usepackage{amsfonts}
\usepackage{enumitem}
\usepackage{float}
\usetikzlibrary{shapes,arrows}

\tikzstyle{obsable} = [draw, ellipse, text width=2.8em, 
    text centered, minimum height=2.4em, fill=black!30]
\tikzstyle{line} = [draw,  very thick, color=black!50, -latex']
\tikzstyle{param} = [draw, ellipse, text centered, text width=2.8em,
    minimum height=2.4em]
\tikzstyle{plate} = [draw, rectangle, text width=6.2cm, 
    text centered, minimum height=15em]
\tikzstyle{ghost} = [draw=none,ellipse, text width=2.8em,
    minimum height=2.4em]
    
\newcommand{\Kep}{\emph{Kepler} }

\newcommand{\Rearth}{R$_\oplus$}
\newcommand{\Mearth}{M$_\oplus$}

\slugcomment{Submitted to the Astrophysical Journal on April 27, 2015}
\shorttitle{Scatter in the Sub-Neptune Mass-Radius Relationship}
\shortauthors{Wolfgang, Rogers \& Ford}

\begin{document}

\title{Probabilistic Mass-Radius Relationship for Sub-Neptune-Sized Planets}

\author{Angie Wolfgang\altaffilmark{1,2,6,7}, Leslie A. Rogers\altaffilmark{3,4,5}, and Eric B. Ford\altaffilmark{6,7,8}}
\altaffiltext{1}{Department of Astronomy and Astrophysics, University of California, Santa Cruz, CA 95064, USA}
\altaffiltext{2}{NSF Graduate Research Fellow; NSF Astronomy \& Astrophysics Postdoctoral Fellow}
\altaffiltext{3}{Department of Astronomy and Division of Geological and Planetary Sciences, California Institute of Technology, MC249-17, 1200 East California Boulevard, Pasadena, CA 91125, USA}
\altaffiltext{4}{Department of Astronomy, University of California Berkeley, 501 Campbell Hall \#3411, Berkeley, CA 94720, USA}
\altaffiltext{5}{Hubble Fellow; Sagan Fellow}
\altaffiltext{6}{Center for Exoplanets and Habitable Worlds, 525 Davey Laboratory, The Pennsylvania State University, University Park, PA 16802, USA}
\altaffiltext{7}{Department of Astronomy and Astrophysics, The Pennsylvania State University, 525 Davey Laboratory, University Park, PA 16802, USA}
\altaffiltext{8}{Center for Astrostatistics, 417 Thomas Building, The Pennsylvania State University, University Park, PA 16802, USA}
\email{akw5014@psu.edu}

\begin{abstract}

The Kepler Mission has discovered thousands of planets with radii $<4$ \Rearth, paving the way for the first statistical studies of the dynamics, formation, and evolution of these sub-Neptunes and super-Earths.  Planetary masses are an important physical property for these studies, and yet the vast majority of \Kep planet candidates do not have theirs measured.  A key concern is therefore how to map the measured radii to mass estimates in this Earth-to-Neptune size range where there are no Solar System analogs.  Previous works have derived deterministic, one-to-one relationships between radius and mass.  However, if these planets span a range of compositions as expected, then an intrinsic scatter about this relationship must exist in the population.  Here we present the first probabilistic mass-radius relationship (M-R relation) evaluated within a Bayesian framework, which both quantifies this intrinsic dispersion and the uncertainties on the M-R relation parameters.  We analyze how the results depend on the radius range of the sample, and on how the masses were measured.  Assuming that the M-R relation can be described as a power law with a dispersion that is constant and normally distributed, we find that $M/M_\oplus=2.7(R/R_\oplus)^{1.3}$, a scatter in mass of $1.9M_\oplus$, and a mass constraint to physically plausible densities, is the ``best-fit" probabilistic M-R relation for the sample of RV-measured transiting sub-Neptunes ($R_{pl}<4$ \Rearth).  More broadly, this work provides a framework for further analyses of the M-R relation and its probable dependencies on period and stellar properties.

\end{abstract}


\section{Introduction} \label{intro}

The \emph{Kepler Mission} has found thousands of planetary candidates with sizes between that of Earth and Neptune \citep{Mul15,Rowe15,Bur14,Bat13,Boru11}.  The emergence of this population poses fundamental questions about the typical compositional constituents of planets within a few times Earth's size.  As bulk densities offer some insight into this problem, mass and radius measurements of individual planets have provided observational constraints for theoretical composition studies performed on a per-planet basis \citep[e.g.][]{Val10,Rog10,Lop12}.  Recently these studies have shifted to considering the available planets as a statistical ensemble (e.g. \citealt{Rog15}; \citealt{Wol15} sans mass constraints), which motivates detailed analyses of the observed mass-radius distribution.

The joint planetary mass-radius distribution, which is often couched in terms of the mass-radius ``relationship" (M-R relation), is also highly relevant for dynamical and formation studies of the \Kep planet candidates (PCs).  Mass measurements for individual PCs are often unavailable, as the majority orbit stars too faint for Doppler follow-up \citep{Bat10} and only $\sim 6\%$ exhibit transit timing variations (TTVs) at high signal-to-noise ratios \citep{Maz13}.  Therefore, a statistical ``conversion" is necessary to map observed radii to the masses these studies need.

To date, several M-R relations have been posed in the exoplanet literature.  To solve the practical issue described above, \citet{Lis11} fit a power law to Earth and Saturn and found $M=R^{2.06}$, where M and R are in Earth units.  \citet{WuY13} derived masses using the amplitudes of sinusoidal TTVs for 22 planet pairs, and found $M=3R$.  More recently, \citet{Wei14}, hitherto WM14, fit a power law to masses and radii available in the literature, which was dominated by the 42 planets chosen by the \Kep team to be followed up with radial velocity measurements \citep{Mar14}; they found $M=2.69R^{0.93}$ for planets with $1.5<R<4$ \Rearth.  

All of these results were produced via basic least squares regression, which is commonly used in astronomy to fit lines through points.  However, this classic technique does not properly account for several issues that are relevant to the small-planet M-R relation: measurement uncertainty in the independent variable (i.e. planet radii), non-detections and upper limits, and intrinsic, astrophysical scatter in the dependent variable (i.e. planet masses).  Thankfully, there are solutions to these problems in both the Bayesian and frequentist statistics literature (see \S1 of \citet{Kel07} for a concise overview).  We present an example of one of these techniques which can be executed using existing numerical algorithms and code (\S\ref{HBM}), which is effectively a simplified implementation of the \citet{Kel07} linear regression scheme.

Of particular interest is the intrinsic scatter that has not been previously characterized.  Theoretical work on planet compositions suggest this scatter should exist: thermally evolved rock-hydrogen sub-Neptune internal structure models yield radii mostly independent of mass \citep{Lop14}, which produces significant mass-radius scatter when a distribution of gaseous mass fractions is present in the population \citep{Wol15}.  Furthermore, the diversity of choices for exoplanets' internal structures produces a range of radii at a given mass due only to differences in the layers' compositions \citep[e.g.][]{Sea07,For07,Rog11}.  

These theoretical findings motivate us to move beyond deterministic, one-to-one mappings, which are in a sense ``mean" relationships.  This average behavior is insufficient and inappropriate if one's aim is to argue for a particular physical process based on full distributions of parameters (versus qualitative comparison to observations), or if the purpose is to rule out parts of parameter space, which requires knowledge of the full mass-radius distribution.  

Realizing the need to move beyond deterministic mass-radius relations for their own theoretical work, \citet{Cha15} derived a piecewise probabilistic M-R relation by fitting the density distribution of planets in four mass bins, and then fit a continuous, yet still deterministic, relation to those results.  However, they stop short of computing a relation which is both continuous and probabilistic (which they admit would be ideal), and do not incorporate measurement error, which is significant for small planets.  With the hierarchical Bayesian modeling that we employ here, we do both.  In the process, we also more fully characterize the uncertainty in the M-R relation based on the current data.  The effort to understand this uncertainty is important, as quantifying how well constrained the M-R relation parameters are will be a key metric by which we measure the improvement in our understanding of the M-R distribution, especially as TESS and its follow-up observations produce more individual mass and radius measurements.

In this paper we show how a probabilistic M-R relation can be constructed (\S\ref{models}) and constrained (\S\ref{HBM}) using any subset of planetary masses and radii (\S\ref{data}).  We also highlight the observational evidence for this expected intrinsic scatter and quantify it in a statistically robust way that includes uncertainties on the M-R relation parameters (\S\ref{res}).  We discuss the correct usage and some major implications of these findings in \S\ref{discuss}.

\section{Modeling the M-R Relation} \label{models}

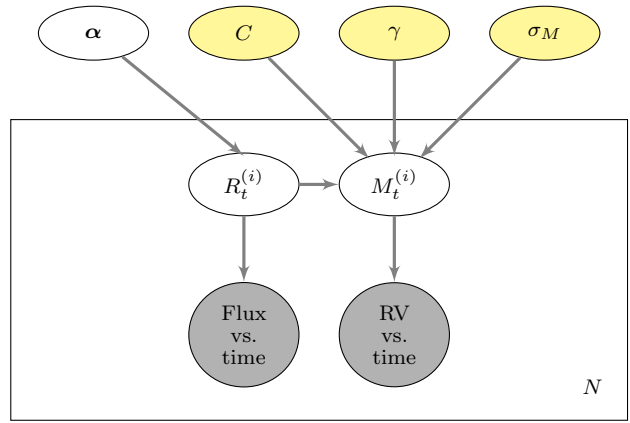
\begin{figure}[t]
\begin{center}
\begin{tikzpicture}[node distance = 2cm, auto]
    \node [param, fill=yellow!50] (gam) {$\gamma$};
    \node [param, left of=gam, node distance = 2cm, fill=yellow!50] (const) {$C$};
    \node [param, right of=gam, node distance = 2cm, fill=yellow!50] (sigM) {$\sigma_M$};
    \node [param, left of=const, node distance = 2cm] (rhyp) {$\bm{\alpha}$};
    \node [plate, below of=const, node distance = 3.15cm, xshift=1cm, minimum height=4cm, text width=8cm] (individMR) {
        \begin{tikzpicture}[anchor=center, node distance = 2cm, auto]
            \node [param] (mtrue) {$M_t^{(i)}$};
            \node [param, left of=mtrue, node distance = 2cm] (rtrue) {$R_t^{(i)}$};
            \node [obsable, below of=mtrue] (mobs) {RV vs. time}; 
            \node [obsable, below of=rtrue] (robs) {Flux vs. time}; 
            \path [line] (rtrue) -- (mtrue);
            \path [line] (rtrue) -- (robs);
            \path [line] (mtrue) -- (mobs);
        \end{tikzpicture}
    };
    \node [ghost, below of=gam, node distance = 2.0cm] (mtrueg){};
    \node [ghost, below of=const, node distance = 2.0cm] (rtrueg){};
    \node [ghost, below of=sigM, text width=0.0cm, minimum height=0.0cm, node distance = 4.7cm, xshift=0.5cm] {$N$};
    \path [line] (gam) -- (mtrueg);
    \path [line] (sigM) -- (mtrueg);
    \path [line] (const) -- (mtrueg);
    \path [line] (rhyp) -- (rtrueg.north);
\end{tikzpicture}
\caption[blah]{
Graphical model used to find the best-fit parameters for the probabilistic mass-radius relationship in Eqn \ref{MR3}.  These parameters of interest are yellow while the observed data are gray (see \S\ref{data}) and unobserved parameters are white; definitions are below.  In practice, we summarize each planet's ``RV vs. time" dataset as the mass measurement $M_{ob}^{(i)}$ and the uncertainty in that measurement, $\sigma_{Mob}^{(i)}$; similarly, the ``Flux vs. time" dataset is summarized as the radius measurement $R_t^{(i)}$ and its uncertainty $\sigma_{Rob}^{(i)}$.  \S \ref{caveats} contains further discussion of this choice.  
Our full hierarchical model, which includes the details of the probability distributions from which each parameter is drawn, is displayed in Equation \ref{statmod}.
    \begin{itemize}[noitemsep, label={}]
       \item $\bm{\alpha}$ = population-wide radius distribution parameters
       \item $C$ = constant in mean M-R relation
       \item $\gamma$ = power law index of mean M-R relation
       \item $\sigma_M$ = intrinsic dispersion in planet masses at a given radius
       \item $R_t^{(i)}$ = true radius of the ith planet
       \item $R_{ob}^{(i)}$ = observed radius of the ith planet
       \item $\sigma_{Rob}^{(i)}$ = measurement uncertainty in $R_{ob}^{(i)}$
       \item $M_t^{(i)}$ = true mass of the ith planet
       \item $M_{ob}^{(i)}$ = observed mass of the ith planet
       \item $\sigma_{Mob}^{(i)}$ = measurement uncertainty in $M_{ob}^{(i)}$
    \end{itemize}
} \label{MRstruct}
\end{center}
\end{figure}

Power laws are often used to parameterize the M-R relation because they are conceptually and computationally simple and can be easily fit to data using the familiar tool of linear regression.  We continue with this choice to facilitate more direct comparisons with previous work and to illustrate how a hierarchical framework enables straightforward extensions to entire families of M-R relations.  In addition, we cast this in terms of $M(R)$ instead of $R(M)$ to address the practical problem of estimating masses from \Kep radii.

In particular, we consider three power law-based M-R relations (Eqns \ref{MR1}-\ref{MR5}).  The first is the form used by most prior studies (see \S\ref{intro}):

\begin{equation}\label{MR1}
\frac{M}{M_\oplus}=C\Big(\frac{R}{R_\oplus}\Big)^\gamma 
\end{equation}
where $M$ is the mass of the planet, $R$ is the planetary radius, and $C$ and $\gamma$ are the parameters to be fit to the data.  This relation is deterministic in the sense that only one mass is allowed for a given radius.  

If instead we want to allow for a range --- that is, if we want to incorporate the expected intrinsic scatter --- then we need to create an M-R relation which specifies how those masses should be distributed at a given input radius.  Again, taking the most simple, familiar, and analytically tractable approach, we choose a Gaussian distribution, where the mean population mass $\mu$ is given by the above power-law relation and where the standard deviation $\sigma_M$ (units of M$_\oplus$) parameterizes the intrinsic scatter in planet masses:

\begin{equation}\label{MR3}
\frac{M}{M_\oplus}\sim\text{Normal}\Big(\mu=C\Big(\frac{R}{R_\oplus}\Big)^\gamma,\sigma=\sigma_M\Big)
\end{equation}
Note that $\sim$ means ``drawn from the distribution", thereby marking the difference between a deterministic and a probabilistic M-R relation.  Figure \ref{MRstruct} is the graphical model corresponding to Eqn \ref{MR3}, and includes Gaussian error bars on the measured masses and radii (see \S \ref{HBM} for all details of the model).

Generalizing further, the width of the intrinsic scatter may change as planets increase in size, so we consider a probabilistic M-R relation that allows the standard deviation itself to vary as a function of radius via the slope $\beta$ (units of M$_\oplus^2$R$_\oplus^{-1}$):

\begin{equation}\label{MR5}
\frac{M}{M_\oplus}\sim\text{Normal}\Big(\mu=C\Big(\frac{R}{R_\oplus}\Big)^{\gamma},\sigma=\sqrt{\sigma_{M1}^2+\beta\tilde{R}}\Big)
\end{equation}
where $\tilde{R}=R/R_\oplus-1$ and $\sigma_{M1}$ is now the standard deviation in planet masses at 1 R$_\oplus$ ($\tilde{R}=0$).

\section{Data} \label{data}

With the statistical M-R relations defined, we turn to the problem of identifying which observational dataset to use.  Optimally we would use a subset of mass and radius measurements that is uniform and complete, as any systematic biases present in the sample will manifest as biased M-R parameter values.  Unfortunately, the available masses and radii are far from this ideal, with mass measurements made with two fundamentally different methods by many different pipelines and chosen for follow-up by a complex, poorly documented selection function.  There is significant work to be done to understand how these systematics affect the M-R relation, but it is outside the scope of this paper, as our main purpose is to show how a probabilistic M-R relation can be derived from whichever dataset one wishes to use.  Therefore, we choose a baseline dataset consisting of radial velocity-measured masses, which somewhat reduces the heterogeneity of the sample while preserving a fairly large number of data points.  

Table \ref{dataset} shows all of the masses and radii that we consider, with our baseline dataset denoted with a label of 0; the list was constructed by starting with the WM14 dataset and identifying new planets and updates in the NASA Exoplanet Archive (last accessed 1/30/2015).  We manually double-checked each planet to verify that the reported measurements were correct and most up-to-date, paying particular attention to which methods and stellar parameters were used (data denoted by a label of 1 were present in and haven't changed since WM14). Given the above concerns with dataset heterogeneity, when both TTV and RV masses are independently available for a single planet, we choose the RV-measured masses.  In practice, only Kepler-18b \citep{Coc11} provide strong enough mass constraints from both methods to require a choice to be made, and even then the two mass measurements are consistent.  The TTV dataset (label of 2) contains only the sub-Neptune-sized planets that have had their transit timing variations fit with N-body integrations, as these masses are the best constrained and therefore provide the most information for the sub-Neptune M-R relation; neither circumbinary planets nor unconfirmed planets 
were included, again to try to keep a somewhat more homogeneous dataset.  
 Finally, to enable easier comparison with previous work, we continued the error treatment of WM14: if asymmetric upper and lower uncertainties were reported, we used their average as a symmetric $1 \sigma$ error bar\footnote{Future work using HBM can improve on this error treatment by using the full posteriors of the mass and radius measurements, if these posteriors are made available in the literature (see \S \ref{caveats}).}.  $2\sigma$ upper limits were included if they were $<80$ \Mearth\ for $R<4$ \Rearth\ and $<300$ \Mearth\ for $4<R<8$ \Rearth.

\section{Fitting the M-R Relations}\label{HBM}

We use hierarchical Bayesian modeling (HBM) to fit the M-R relations in \S\ref{models} to the data described in \S\ref{data}.  This statistical method is described in detail in \citet{Wol15} in the context of exoplanet compositions; further pedagogical discussion and examples of HBM in the astronomical literature is provided by \citet{Lor13}.  A very similar approach to this HBM-enabled linear regression was detailed in \citet{Kel07}; we refer the reader to that paper for an in-depth discussion of the general advantages and improvements of this approach over the commonly used $\chi^2$ analysis for linear regression.

For the problem at hand, HBM (or the analogous frequentist methods for multi-level modeling) is necessary for a number of reasons:
\begin{itemize} 
  \item It allows us to directly model and fit the astrophysical dispersion in the population as an explicit parameter.
  \item It allows us to self-consistently incorporate uncertainties on the independent variable (radii in this case), without the need for elaborate bootstrapping schemes.
  \item Most sub-Neptune mass uncertainties are large, and some are realistically only upper limits.  HBM is able to simultaneously use all likelihood distributions no matter their width or shape, which increases the information content of the resulting M-R relation and decreases the biases that binning or weighting schemes introduce when these likelihoods are asymmetric.
  \item Relatedly, HBM allows us to introduce the true masses and radii as latent (unobserved) parameters; this enables us to restrict the masses to physically allowed parameter space (such as $M>0$ or $\rho<\rho_{iron}(M)$) while preserving all of the information in the observations, including when the ``best-fit" masses happen to be negative.
  \item As with all Bayesian methods, HBM produces posterior distributions, allowing us to easily see the uncertainties in the M-R relation parameters.  Most of the M-R relations currently reported and used in the literature have no published uncertainties.
 \end{itemize}
 
 \begin{figure*}[t]
\begin{center}
\includegraphics[angle=270,scale=0.65]{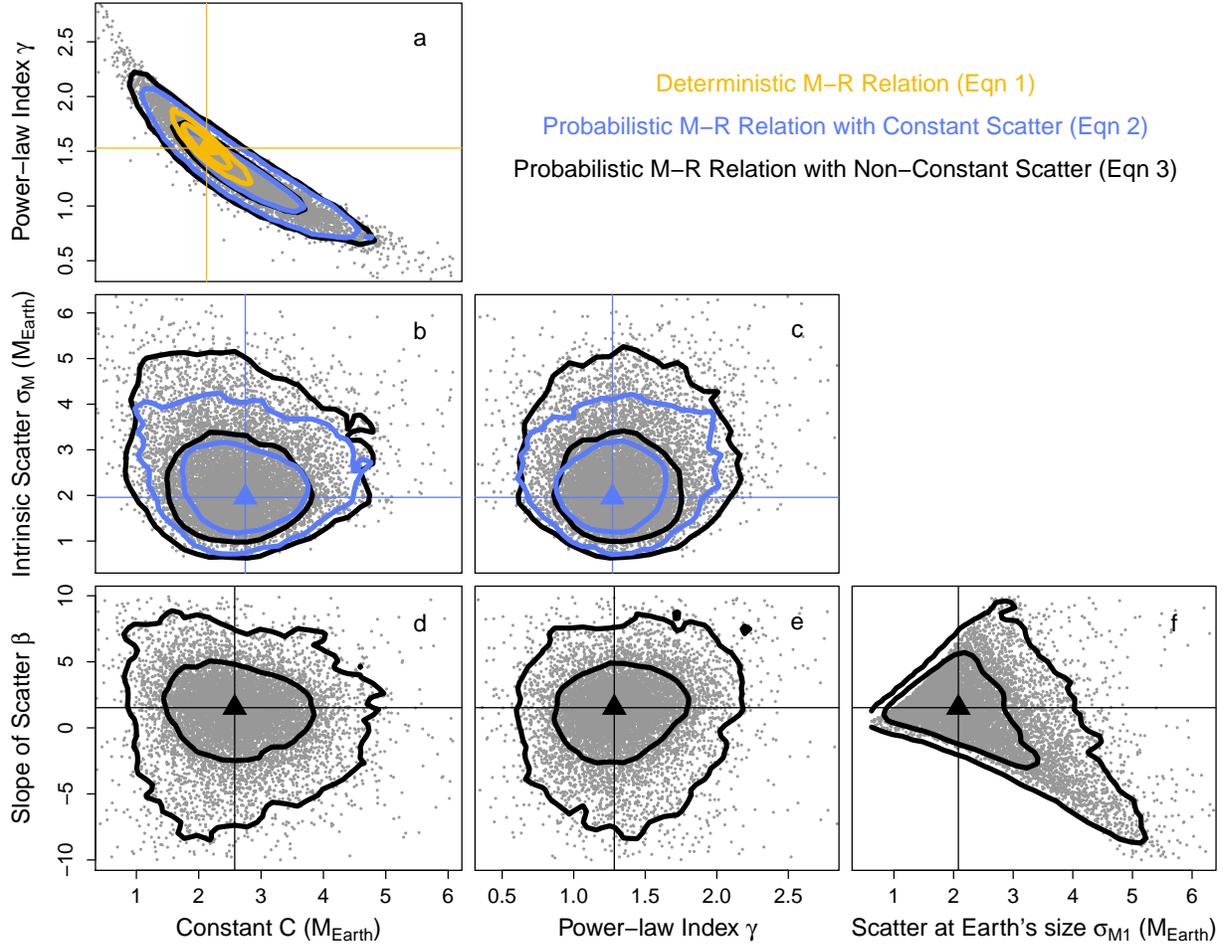}
\caption{Posteriors for the parameters in our family of M-R relations (row 1: Equations \ref{MR1}-\ref{MR5}; row 2: Equations \ref{MR3}-\ref{MR5}; row 3 and gray posterior samples in all panels, which contain all 10,000 saved samples from our thinned MCMC chains: Equation \ref{MR5}) when fit to our baseline dataset.  68\% and 95\% contours are shown for each, and demarcate the uncertainties on these M-R relation parameters; the triangles denote best-fit values.  Panels b and c show that $\sigma_M=0$ is strongly excluded for $R<4$ \Rearth, and so astrophysical scatter exists in the sub-Neptune M-R relation.  Therefore, theoretical studies which require an M-R relation should use a probabilistic one like that of Eqn \ref{MR3} with one of the sets of parameter values in Table \ref{MRres}.  
} \label{compMRs}
\end{center}
\end{figure*}

The hierarchical model for our baseline M-R relation (Eqn \ref{MR3}) is displayed in Figure \ref{MRstruct} to clarify the structural relationships between parameters and observables.  This structure is also present in the written version below, along with details of the distributions we used (``N" represents a normal distribution with the listed parameters in order of $\mu$ and $\sigma$; ``U" represents a uniform distribution with the listed numbers bounding the interval; and ``$|$" means ``given", i.e. the parameter to the left depends on the parameters to the right):
\begin{align}\label{statmod}
\gamma&\sim\text{N}(1,1)\nonumber\\
\text{ln}(C)&\sim\text{U}(-3,3)\nonumber\\
\text{log}(\sigma_M^2)&\sim\text{U}(-4,2)\nonumber\\
R_t^{(i)}&\sim\text{U}(\alpha_1=0.1,\alpha_2=10)\nonumber\\
\mu_M^{(i)}|R_t^{(i)},C,\gamma&=\gamma\text{ln}(R_t^{(i)})+\text{ln}(C)\nonumber\\
M_t^{(i)}|R_t^{(i)},C,\gamma,\sigma_M&\sim\text{N}\Big(\mathrm{e}^{\mu_M^{(i)}},\sigma_M\Big)\nonumber\\
R_{ob}^{(i)}|R_t^{(i)},\sigma_{Rob}^{(i)}&\sim\text{N}(R_t^{(i)},\sigma_{Rob}^{(i)})\nonumber\\
M_{ob}^{(i)}|M_t^{(i)},\sigma_{Mob}^{(i)},R_t^{(i)},C,\gamma,\sigma_M&\sim\text{N}(M_t^{(i)},\sigma_{Mob}^{(i)}) 
\end{align}
For the deterministic M-R relation of Eqn \ref{MR1}, Eqn \ref{statmod} remains the same except there is no $\sigma_M$ parameter, and 
\begin{align}
M_t^{(i)}|R_t^{(i)},C,\gamma&=\mathrm{e}^{\mu_M^{(i)}}\nonumber
\end{align}
while for the M-R relation of Eqn \ref{MR5}, there was an additional parameter $\beta$ such that:
\begin{align}
\beta&\sim\text{U}(-10,10)\nonumber\\
M_t^{(i)}|R_t^{(i)},C,\gamma,\sigma_{M1},\beta&\sim\text{N}\Big(\mathrm{e}^{\mu_M^{(i)}},\sqrt{\sigma_{M1}^2+\beta\tilde{R}_t^{(i)}}\Big)\nonumber
\end{align}
Note that the normal distributions in the last two lines of the model (collectively Eqn \ref{statmod}) are the same likelihoods that are assumed when using $\chi^2$ to perform linear regression.

For all M-R relations we consider, we truncated the $M_t^{(i)}$ distribution such that $0<M_t^{(i)}< M_{t,pureFe}^{(i)}$ where $M_{t,pureFe}^{(i)}$ was computed using the 0\% rock mass fraction analytic fits to the \citet{For07} rock-iron internal structure models:  
\begin{equation}\label{maxM}
log(M_{t,pureFe}^{(i)})=\frac{-b + \sqrt{b^2-4a(c-R_t^{(i)})}}{2a}
\end{equation}
where a = 0.0975, b= 0.4938, and c=0.7932 \citep{For07erat}.  This truncation was imposed to restrict the planet masses to physically plausible densities given the planet's true radius.

We also performed a prior sensitivity analysis on our population-wide parameters to assess the degree to which our results depend on our priors.  As given in Eqn \ref{statmod}, we used a wide normal distribution for the power law index; we chose this prior to be wide with a variance of 1 to minimize the amount of information in the posterior that comes from the prior, and we chose a mean of 1 because we expected {\it a priori} that near-linear relationships were physically plausible.  This choice was admittedly arbitrary, so we tested the sensitivity of our results to these assumptions.  To do this, we ran the MCMC again, but with two other priors: a uniform distribution for $\gamma$ and a uniform distribution for $s$ where $\gamma =$ tan($s$), which corresponds to a uniform distribution in the slope of the power law\footnote{Uniform $\gamma$ places high probability at steep power laws, which are highly unlikely on physical grounds.}.  Under each prior, we find that the posterior modes (the ``best fits") for $C,\gamma$, and $\sigma_M$ differed by no more than 0.05 from the best-fits listed in Table \ref{MRres}, which is below our reported precision.  Additionally, we tested several end member cases for the $R_t$ distribution, and the choice for this prior had a similarly negligible effect on the result, primarily because $R_{ob}$ is fairly well constrained throughout the sample.  Therefore, we conclude that our results are robust to our choice of priors.

\begin{figure*}[t]
\begin{center}
\includegraphics[angle=270,scale=0.65]{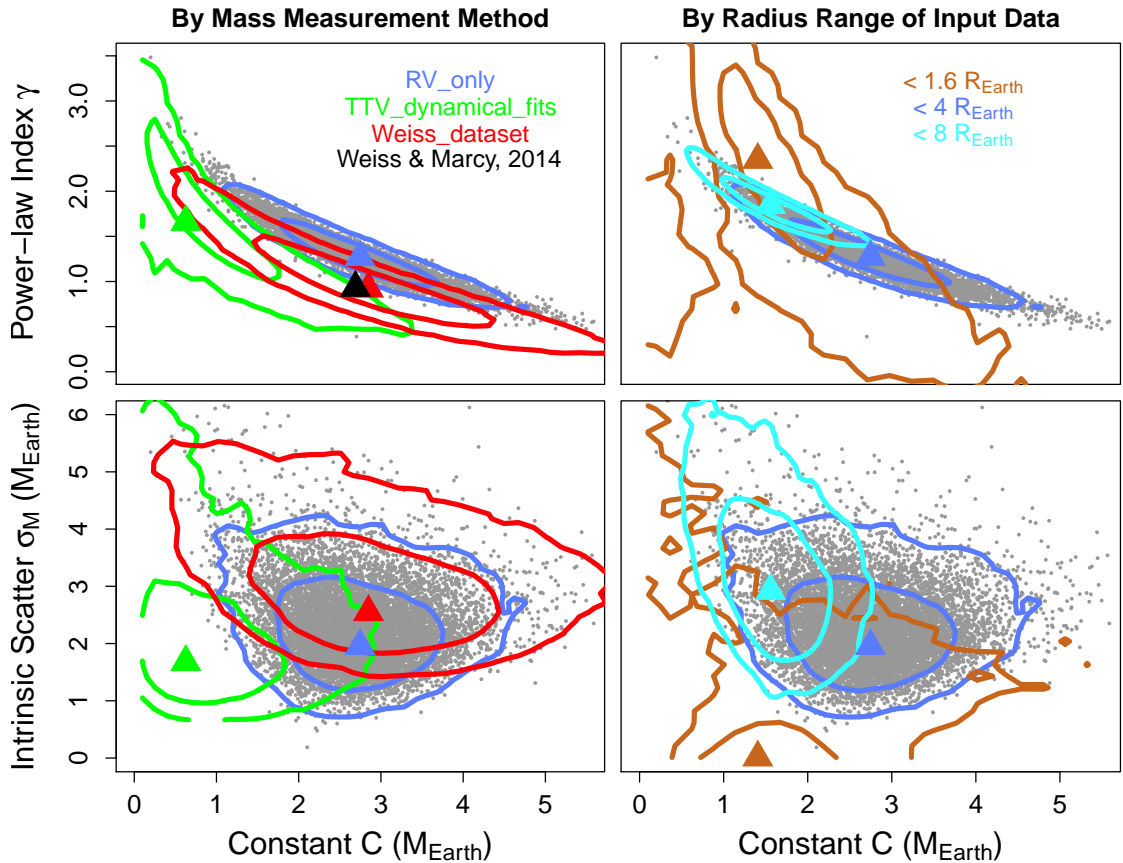}
\caption{Posteriors for Eqn \ref{MR3}'s M-R relation parameters when we change the input dataset (68\% and 95\% contours shown; triangles are best-fit values).  The blue contours represent the baseline dataset and are the same as those in panels a and b of Figure \ref{compMRs}; the gray points are the 10,000 saved posterior samples from our MCMC chains using this baseline dataset.  The green TTV M-R relation is systematically shifted downward (lower $C$) compared to the baseline M-R relation, while the red WM14 dataset, a hybrid of the two, produces a posterior which falls between them (the black point is the WM14 result itself).  When we consider different radius ranges, we see that $R_{obs}<8$ \Rearth\ (cyan) produces a slightly down-shifted, steeper, and more dispersed M-R relation than the default $R_{obs}<4$ \Rearth\ (lower $C$ and higher $\gamma,\sigma_M$, although the posteriors do overlap), while the M-R relation for $R_{obs}<1.6$ \Rearth\ (orange) is not well constrained (although $\sigma_M\approx0$ for reasonable values of $C$).} \label{compdatasets}
\end{center}
\end{figure*}

To produce the results shown in \S\ref{res}, we evaluate each model with JAGS (Just Another Gibbs Sampler; \citealt{Plu03}), an R code for numerically evaluating hierarchical Bayesian models with MCMC.  For each set of posteriors in Figures \ref{compMRs} and \ref{compdatasets}, we ran 10 chains consisting of 500,000 iterations each.  The first half of each chain is discarded as ``burn-in", and the resulting half is thinned by a factor of 250, such that we retain 10,000 posteriors samples of each parameter.  

To assess the independence of these samples, we compute the effective sample size (ESS), which accounts for the autocorrelation still present within these thinned Markov chains (ESS $=10000$ indicates perfect independence).  The ESS is $>4000$ for each parameter listed in Table \ref{MRres}, with two exceptions.  The ESS of the deterministic relation parameters are around 230, an order of magnitude lower than all the probabilistic relations we tested.  The difficulty this model had with convergence reflects the challenges of applying a physically inappropriate model to data and is another indication that a deterministic relation does not fit the observed masses and radii well.  Less concerning yet not quite as well converged compared to the others were the parameters for the probabilistic M-R relation fit only to the smallest radii (ESS $= 1500-4000$).  This occurs because these small planets have the largest mass uncertainties; this causes the maximum mass restriction in Eqn \ref{maxM} to severely truncate most of the likelihoods, which results in high autocorrelation in the MCMC chains.  Given the small ESS for these two sets of parameters, we caution against over-interpretation of their results: their ``best-fit" values in Table \ref{MRres} are $\sim 6$ and $\sim 2$ times more uncertain than the others (corresponding to the precision of the posterior mean with the square root of the sample size [the ESS]), and the boundaries of their 95\% credible regions in Figure \ref{compdatasets} are poorly estimated.  We do not spend time running these simulations longer, as they were performed for the sake of comparison, and we display them for this purpose only.  

For the main result --- the baseline dataset evaluated with the probabilistic M-R relation --- the ESS of $C$ and $\gamma$ are 10,000, and the ESS of $\sigma_M$ is 5,300.  Furthermore, the between-chain convergence diagnostic $\hat{R}$ of \citet{Gel92} is $\leq1.002$ for all parameters in our probabilistic models (except again for the smallest radii planets, for which $\hat{R} = 1.008$ at its worst).  Together, these two tests provide no evidence that the posteriors have not converged, and we proceed with the usual amount of confidence (given that no one can ever prove convergence).

\section{Results}\label{res}

\begin{deluxetable}{cccccc}
\centering
\tabletypesize{\footnotesize}
\tablecolumns{6}
\tablewidth{0pt}
\tablecaption{Best-Fit Parameters of the M-R Relation \label{MRres}}
\tablehead{
\colhead{Equation} & \colhead{Dataset} & \colhead{$C$}  & \colhead{$\gamma$}  & \colhead{$\sigma_M$}  & \colhead{$\beta$}
}
\startdata
1 & baseline: RV only, $< 4$ \Rearth & 2.1 & 1.5 & --- & --- \\
2 & baseline: RV only, $< 4$ \Rearth & 2.7 & 1.3 & 1.9 & --- \\
2 & N-body TTVs only, $< 4$ \Rearth & 0.6 & 1.7 & 1.7 & --- \\
2 & Weiss ($< 4$ \Rearth) & 2.8 & 0.9 & 2.5 & --- \\
2 & RV only, $< 1.6$ \Rearth & 1.4 & 2.3 & 0.0 & --- \\
2 & RV only, $< 8$ \Rearth & 1.6 & 1.8 & 2.9 & --- \\
3 & baseline: RV only, $< 4$ \Rearth & 2.6 & 1.3 & 2.1 & 1.5 
\enddata
\tablecomments{These ``best fit" values correspond to the mode of the joint posterior distributions; see code and posterior samples in the github repository dawolfgang/MRrelation to account for the full uncertainty in the parameters that is contained the posteriors (see \S \ref{visualize} for more details on this).  Also, when using these M-R relations to generate masses from planet radii, one should apply the density constraint given by Eqn \ref{maxM}.}
\end{deluxetable}

Table \ref{MRres} shows the results of our modeling: it displays the best-fit parameters for each of the various M-R relations and datasets that we consider.  The first entry corresponds to the deterministic M-R relation; entries $2-6$ correspond to our probabilistic M-R relation for various datasets (see \S \ref{data}, \S\ref{diffdat}); and the last entry corresponds to the probabilistic M-R relation with non-constant scatter.  In particular, the second entry lists the best-fit values for Eqn \ref{MR3} using our baseline dataset.  All were computed with the density restriction given by Eqn \ref{maxM}; this constraint should also be applied to the masses generated from these M-R relations when these relations are used in forward modeling.

In all cases the reported ``best fit" values correspond to the mode of the joint posterior distribution, and are denoted by the triangles in Figures \ref{compMRs}-\ref{compdatasets}.  The uncertainties in the parameters are represented by the displayed 68\% and 95\% posterior contours, with the contours corresponding to our baseline dataset colored blue.  The gray points are the 10,000 saved posterior samples from our thinned MCMC chains using the baseline dataset.

\subsection{Deterministic vs. Probabilistic M-R Relations}\label{defaultMR}

The primary motivation for this paper was to assess the observational evidence for intrinsic scatter in the sub-Neptune M-R relation, and to characterize this scatter if warranted.  To do so, we compare the posteriors for our three M-R relations in Figure \ref{compMRs} (note that not all relations have all parameters: for example, the deterministic M-R relation of Eqn \ref{MR1} is described only by $C$ and $\gamma$, so it only appears in panel a).  Panels b and c show that this intrinsic scatter exists: because the posteriors lie away from zero, $\sigma_M=0$ is strongly excluded by the data, even with the currently large individual mass error bars.  This is not a result of our choice of priors: the parameterization in Eqn \ref{statmod} is equivalent to $\sigma_M^2\sim1/\sigma_M^2$, which is strongly weighted toward zero, in contrast to the posterior we compute.

Comparing the different M-R relations, we see that the $C,\gamma$ posterior for the model given by Eqn \ref{MR1} is much tighter than that for Eqns \ref{MR3}-\ref{MR5}.  This is expected: when we keep the dataset fixed but add more parameters, especially one like $\sigma_M$ that by construction allows wiggle room around a deterministic relation, the observational information content per parameter decreases, and the posteriors widen.  Given this expectation, what is arguably more notable are the small differences between Eqn \ref{MR3} and \ref{MR5}'s model posteriors for the parameters they have in common: most of the extra width of Eqn \ref{MR5}'s joint posterior is contained in the new parameter $\beta$ (Figure \ref{compMRs}, panels d-f), which spans zero.  There is therefore not enough evidence in the current dataset to justify an intrinsic scatter that changes as a function of radius, at least not under our model assumptions\footnote{While outside the scope of this paper, future analyses of the M-R relation can address this and other questions of model selection more quantitatively by computing posterior Bayes factors.  Regardless, the results for the statistical models represented by Eqns \ref{MR1} and \ref{MR5} can serve as a sensitivity test for that of Eqn \ref{MR3}, as we describe.}.  For the best-fit values of these parameters, which correspond to the triangles in Figure \ref{compMRs}, see Table \ref{MRres}.

\begin{figure*}[t]
\begin{center}
\includegraphics[angle=270,scale=0.65]{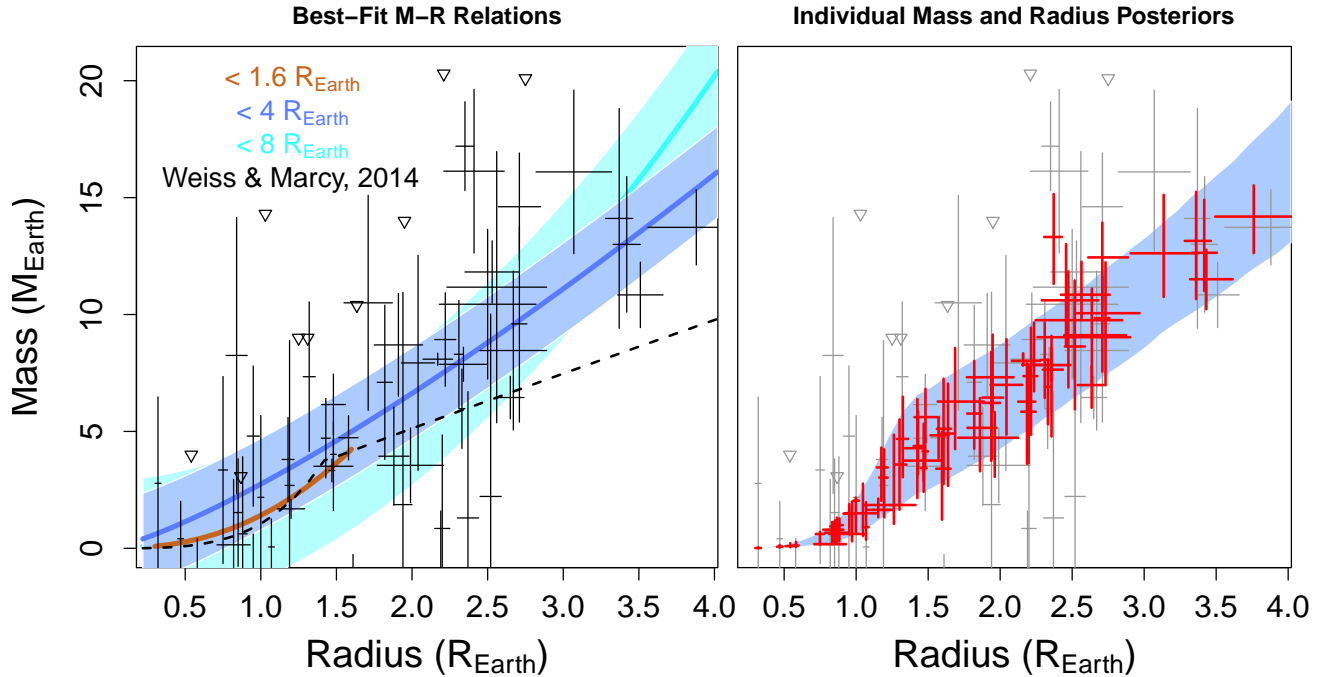}
\caption{\emph{Left:} the best-fit M-R relations from the right column of Figure \ref{compdatasets}.  For each, the solid line denotes the mean relation $\mu_M$ while the faded region denotes the standard deviation of the intrinsic scatter (vertical height of region to either side of line $=\sigma_M$; note $\sigma_M=0$ for the smallest planets).  The M-R relation of WM14 is the dashed black line while the baseline dataset is overplotted as the thin black lines with triangles for the 2-$\sigma$ upper limits (note that WM14 was calculated with a dataset that includes TTV planets).  \emph{Right:} the baseline M-R relation (the second entry in Table \ref{MRres}) marginalized over the corresponding posterior distribution and subjected to our physical mass range restriction.  The blue region now corresponds to the central 68\% of planet masses that were drawn at a given radius. The 68\% coverage interval of the posterior true masses and radii of individual planets are plotted red; the same $R_{ob}$ and $M_{ob}$ as on the left are plotted in gray for comparison.} \label{plotMRs}
\end{center}
\end{figure*}

 \subsection{Changing the Dataset}\label{diffdat}
 
The results in \S\ref{defaultMR} are for our baseline dataset, an RV-only sample with $R_{obs}<4$ \Rearth.  However, all Bayesian results depend on the data that are used, so it is important to carefully consider what the dataset contains.  To demonstrate this, we present some illustrative examples of the M-R relation posteriors under different mass and radius selection functions (Figure \ref{compdatasets}).

The left side of Figure \ref{compdatasets} displays results for samples of planets that have had their masses measured in different ways.  A number of prior studies (e.g. \citealt{Jon14}, WM14) have noted that planets with masses measured from their high SNR TTVs tend to be systematically less dense than planets with RV-measured masses\footnote{This appears to be a population-level effect: there are few planets with independently analyzed, strongly constrained TTV and RV masses, and they all yield measurements that are consistent between the two methods [Kepler-18b \& c \citep{Coc11}, Kepler-88c \citep{Nes13}, and Kepler-117 \citep{Bru15}; note that only Kepler-18b appears in our table]. There are other systems where the two methods have been used in concert to infer planet masses (e.g., Kepler-9, Kepler-10 Kepler-11, Kepler-89; see table for references), but the constraints from one or both methods are relatively weak, making them insensitive tests of a TTV-RV difference on a per-planet basis.}.  The origin of this difference is still unclear, and is not something our current modeling is able to address (see \S \ref{caveats}): it could be due to either an intrinsic difference in the densities of these two populations, or to observational bias, as TTVs for larger (and thus less dense) planets are easier to detect while RVs for more massive (and thus more dense) planets are easier to measure.  In any case, our results confirm the existence and nature of this discrepancy, if not the reason: the green TTV-only posterior is shifted towards lower $C$ with similar $\gamma$ and $\sigma_M$, which produces on average lower masses for a given radius.  Furthermore, the hybrid WM14 dataset yields the red posterior, which falls between the TTV-only and RV-only posteriors yet peaks at lower $\gamma$, illustrating that posterior modes (Bayesian ``best fits") for joint datasets are not necessarily averages of the modes for separate subsets.  This behavior can be understood when one considers that these TTV-measured planets are preferentially larger than the RV-measured planets: this pulls the joint M-R relation down at higher radii because the TTV-measured planets there have lower masses (which lowers $\gamma$) but affects the relation at lower radii very little because there are few small planets in our TTV-measured sample (which keeps $C$ roughly the same).

The right side of Figure \ref{compdatasets} displays results for samples of planets spanning different radius ranges, illustrating the effect that a somewhat arbitrary radius cut can have on one's results.  Compared to our default sub-Neptune range, a $R_{obs}<8$ \Rearth\ cut produces an M-R relation that is overall shifted down, is steeper, and has more intrinsic scatter (the cyan posterior has lower $C$ and higher $\gamma,\sigma_M$).  This is consistent with the \citet{Lis11} fit to Earth and Saturn over a similar radius range, although neither of these Solar System planets were included in our dataset.  Meanwhile, the M-R relation is poorly constrained for the $R_{obs}<1.6$ \Rearth\ sample, the radius range outside of which rocky planets likely do not occur \citep{Rog15}.  This is because our $0<M_t^{(i)}< M_{t,pureFe}^{(i)})$ restriction is most severe for these small planets, allowing only a small range of physically plausible masses.  This range is completely spanned by most of the mass measurements (see right side of Figure \ref{plotMRs}), so there is little empirical extrasolar information for $R_{obs}<1.2$ \Rearth, 
and the orange posteriors are dominated by the few larger planets with well measured masses.  With this sample, there is not currently enough observational evidence in this radius range to rule out a deterministic relation. This does not mean, however, that these small planets all have the same composition, as the posterior 68\% contour spans power-law indexes between 1 and 3 and a constant population-wide Earth-like rocky composition would have a power-law index of around 3.7 \citep{Val06}.  More data will be needed to yield a better estimate of the power-law index, and therefore of the compositional diversity of small planets.

 \subsection{Visualizing the M-R Relation}\label{visualize}
 
While the posterior contours in Figures \ref{compMRs}-\ref{compdatasets} show the best-fit M-R relation parameters and their uncertainties, visualizing the M-R relation itself requires that they be mapped from parameter space to mass, radius space.  There are at least two ways to do this with Bayesian analysis, and they are displayed in Figure \ref{plotMRs}.  

First, one can simply take the best-fit values and plot the resulting relation, as was done in the left panel.  Here the $1 \sigma$ width of the probabilistic relation, as parameterized by $\sigma_M$, is denoted by the faded colored region while the mean relation, as parameterized by $C$ and $\gamma$, is the thick line of the same color.  Note that the mean M-R relations extend into unphysical regimes for $R < 1$ \Rearth; this is because the mass observations span the physically allowed region, as discussed in \S\ref{diffdat}, leaving the M-R relation to be constrained primarily by the locations of the larger, higher mass planets and our model assumptions.  The presence of intrinsic scatter in our M-R relation nevertheless allows physically realistic masses to be assigned to the smallest planets; to force this requirement, we recommend adding a density constraint to Eqn \ref{MR3} such that the probability of a planet being drawn outside this range is 0 (the constraint we used is given in Eqn \ref{maxM}), or to use a different M-R relation for sub-Earth-sized planets. The different colors in the left panel correspond to the M-R relations in the right column of Figure \ref{compdatasets}; these mostly overlap in the sub-Neptune regime.  Note that the RV-only dataset produces a steeper relation than one which also contains high SNR TTV planets (i.e. the black dashed WM14 relation), as discussed in \S\ref{diffdat}.  

While these best-fit M-R relations are easy to use, they do not take into account the fact that the posteriors have non-zero width and therefore a range of M-R relation parameters are allowed by any one dataset.  A more thorough implementation of these results would account for these uncertainties by integrating over all of the posterior samples.  This marginalization, which also incorporates the physical restrictions on $M_t$ as given by Eqn \ref{maxM}, is displayed on the right: now the blue region corresponds to the central 68\% of planet masses that were drawn for a given radius.  Note that this region is wider than that on the left and that the masses no longer extend into unphysical regimes.  The 68\% coverage interval of the posterior true masses and radii of individual planets in the baseline sample are plotted red, while the same $R_{ob}$ and $M_{ob}$ as on the left are plotted in gray.  As expected (see the end of \S \ref{modelcheck}), the posteriors have ``shrunk" toward the mean relation within the uncertainties provided by the data. Furthermore, one can readily see that the data are qualitatively consistent with the modeled M-R relation: the red lines fall within and immediately around the blue region (see \S \ref{modelcheck} for a quantitative treatment of model checking).

\section{Discussion}\label{discuss}

\subsection{Using the M-R Relation to Predict Masses} \label{predM}
 
The most straightforward and computationally simple way to map a sub-Neptune's radius to a mass while accounting for intrinsic scatter is to adopt Eqn \ref{MR3} with one of the sets of parameters in Table \ref{MRres} and impose a density constraint like Eqn \ref{maxM} for the smallest planets.  This best-fit M-R relation is analytic and represents a substantial improvement over the previous deterministic relationships in capturing the full mass-radius distribution.  However, it does not incorporate uncertainties in the M-R relation parameters or uncertainties in the measured planet radius itself.  Depending on how detailed one's analysis needs to be, a more accurate predictive mass distribution may be needed.

To account for these issues, one must compute the posterior predictive M-R relation, which marginalizes over both the posteriors displayed here and the radius posterior produced by one's light curve modeling.  This mass distribution will be wider than that produced by simply applying Eqn \ref{MR3} (see right side of Figure \ref{plotMRs}) because it incorporates the above sources of uncertainty and thus more accurately reflects our state of knowledge about these planets' masses.  Kepler-452 b \citep{Jen15} provides an example of an individual planet's posterior predictive mass distribution that has been calculated with this probabilistic M-R relation; because its computation requires the numerical posterior samples that we have produced, the resulting mass distribution is also numerical in nature.  To enable more calculations like this one, we have posted our posterior samples in the github repository dawolfgang/MRrelation along with R code that uses them to compute and plot the posterior predictive mass distribution for individual planets.

\subsection{Model Checking} \label{modelcheck}

The purpose of the right panel in Figure 4 is to provide a qualitative comparison between the data and the baseline probabilistic M-R relation; this visual check immediately verifies that, in the broadest sense, our model is a reasonable description of the data (see \S \ref{visualize}).   However, no model perfectly describes nature.  A more in-depth look is warranted, to both understand the limitations of the current model and to identify areas for improvement in future work.

Quantitatively, we can check the data-model consistency by computing a ``hierarchical p-value", which yields the fraction of all possible datasets which are more discrepant from the model than the observed dataset.  This calculation necessarily involves sampling from our forward model (Eqn \ref{statmod}) and using those samples to calculate a statistic which quantifies ``discrepant".  The identification of robust and useful hierarchical statistics is still an active area of statistical research (see, for example, \citealt{Bay07}), as the multi-level nature of hierarchical modeling offers a number of choices that test different aspects of the model.  As a result, model checking in practice can be an involved process that requires investigations into multiple parts of the problem.  We provide two such investigations below to illustrate some of the subtlety of this endeavor.

Regardless of the details, most forms of Bayesian model checking use the posterior predictive distribution.  Conceptually, this distribution defines the probability of observing a certain data value given the currently observed dataset and the model.  Mathematically, the posterior predictive distribution for a new observation $x_{new}$ is defined as:
\begin{equation}\label{postpredeq}
p(x_{new}|\bm{\hat{x}},M_1) = \int p(x_{new}|\bm{\theta},M_1) p(\bm{\theta}|\bm{\hat{x}},M_1) d\bm{\theta}  
\end{equation}
where $p(x_{new}|\bm{\theta},M_1)$ is the likelihood of a new, currently unobserved data point given the parameters $\bm{\theta}$ of model $M_1$, and $p(\bm{\theta}|\bm{\hat{x}},M_1)$ is the posterior of model $M_1$, i.e. the joint probability of all of $M_1$'s parameters given the observed dataset $\bm{\hat{x}}$.  The integral denotes the process of marginalization over the parameters, so that the probability of a new observation incorporates the model uncertainty allowed by the current observations.  

\begin{figure}[t]
\begin{center}
\includegraphics[angle=270,scale=0.4]{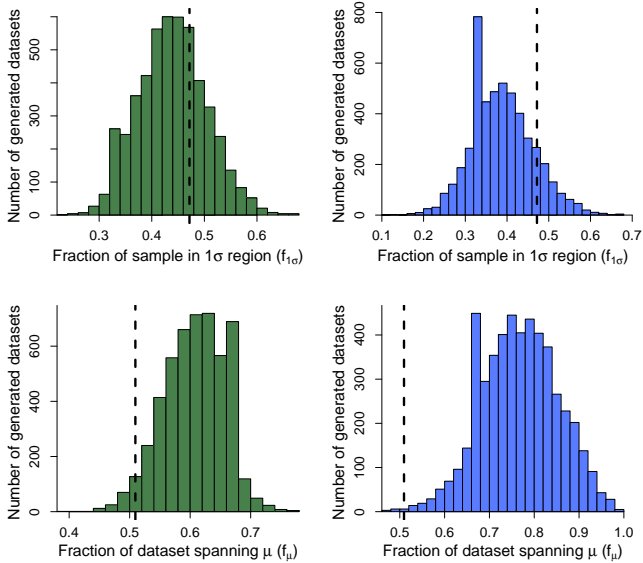}
\caption{Checking the model against the observed data, using two different statistics to quantify the data-model consistency (see \S \ref{modelcheck}).  All panels contain 5,000 hypothetical datasets generated from the baseline posterior predictive distribution (Eqns \ref{postpredeq}-\ref{forwardmod}), with the statistic for the observed dataset (Table \ref{dataset}) as the vertical black dashed line. The green histograms on the left show the consistency of the model with the current dataset, while the blue histograms on the right show the consistency with new hypothetical datasets.  The top histograms show the distribution of the statistic $f_{1\sigma}$.  The observed $f_{1\sigma}$ falls at the 63rd and 83rd percentiles of the distribution (left and right, respectively), indicating that the width of the model's M-R relation is consistent with the data.  The bottom histograms show the distribution of the statistic $f_\mu$. The observed $f_\mu$ falls at the 2nd and 0.2th percentiles of the distribution, indicating that there is room for future improvement on the choice of what distribution to use for the probabilistic M-R relation.} \label{postpred}
\end{center}
\end{figure}

Next, one draws hypothetical data from this posterior predictive distribution until a dataset of the same size N as the observed dataset is achieved:
\begin{equation}\label{drawdataset}
\bm{\hat{x}_{new}} \underset{N}{\sim} p(x_{new}|\bm{x},M_1)
\end{equation}
where $\bm{\hat{x}_{new}} = \left \{ \hat{x}_{new,1},\hat{x}_{new,2},...,\hat{x}_{new,N} \right \}$ and ``$\underset{N}{\sim}$" means ``draw from that same distribution N times".  This dataset is then used to compute the statistic of choice (see discussion below).  Repeating this process thousands of times (i.e. bootstrapping the statistic) generates a distribution of the statistic which can then be compared to the value that was calculated from the observed dataset.  If the observed statistic falls within this distribution, the model is consistent with the data (see Figure \ref{postpred}).

In practice, MCMC simulations (\S \ref{HBM}) provide samples from the posterior rather than the analytic form of the posterior as required in Eqn \ref{postpredeq}; therefore, the posterior predictive distribution is not directly calculated.  Instead, Eqns \ref{postpredeq} and \ref{drawdataset} are combined by using the posterior samples:
\begin{equation}
\bm{\hat{\theta}} \sim p(\bm{\theta}|\bm{\hat{x}},M_1)  \nonumber
\end{equation}
to define the likelihood that one then draws from:
\begin{equation}\label{forwardmod}
\bm{\hat{x}_{new}} \underset{N}{\sim} p(x_{new}|\bm{\hat{\theta}},M_1)   
\end{equation}
Performing these two steps repeatedly produces an ensemble of datasets drawn from Eqn \ref{postpredeq}.  Applying Eqn \ref{forwardmod} to our M-R relation requires evaluating the lower levels of the forward model described in Equation \ref{statmod}.  

Part of the subtlety of checking data-model consistency arises because our model is hierarchical.  In particular, $\bm{\hat{\theta}}$ of Eqn \ref{forwardmod} includes both the population-wide parameters $C,\gamma,\sigma_M$ and the individual parameters $M_t^{(i)},R_t^{(i)}$; we can use the posterior samples of either of these groups of parameters to calculate $\bm{\hat{x}_{new}}$.  We show the result of both choices in Figure \ref{postpred}, using the posterior samples from the baseline M-R relation (second line of Table \ref{MRres}).  Conceptually, using samples from the individual true mass and radius posteriors ($\bm{\hat{\theta}}=\left \{ M_t^{(i)},R_t^{(i)} \right \}$) evaluates the fit of the model to the currently observed set of planets (the green histograms on the left), while using posterior samples for the M-R relation parameters ($\bm{\hat{\theta}}=\left \{ C, \gamma, \sigma_M \right \}$) evaluates the fit of the model to altogether new sets of planets (the blue histograms on the right).  For the former case, the same $\sigma_{Mob}^{(i)}$ and $\sigma_{Rob}^{(i)}$ as the observed dataset are used; for the latter case, $\sigma_{Mob}^{(i)}$ is drawn from the distribution of $\sigma_{Mob}$ for the observed planets that have a similar mass, and $\sigma_{Rob}^{(i)}$ is drawn from the observed dataset's full distribution of $\sigma_{Rob}$ without controlling for radius (the size of the radius error bars are fairly constant across the dataset).

The second aspect of hierarchical model-checking which requires some effort is the identification of a robust statistic to quantify the discrepancy between the observed data and the model-generated data.  Optimally this statistic would test the fit of every part of the modeled probability distribution, including the ``average" behavior of the model, the ``extreme" behavior out on the tails of the distribution, and for hierarchical models, the accuracy of the grouping and relational structure that is illustrated by graphs like Figure \ref{MRstruct}.  Due to high dimensional parameter space, this proves to be very difficult, so one must identify several statistics which test such aspects separately.  To illustrate the problem, we choose two: $f_{1\sigma}$, the fraction of a given dataset's simulated mass, radius observations which fall within the 68\% coverage interval of our probabilistic, baseline M-R relation (blue region in Figure \ref{plotMRs}), and $f_\mu$, the fraction of a given dataset's mass measurements whose $1\sigma$ error bars cross the mean relation $\mu$ (see Eqn \ref{MR3}) of that same model.  Therefore, $f_{1\sigma}$ tests how well the width of our probabilistic M-R relation fits the data, and $f_\mu$ tests how tightly grouped the data are around the mean compared to the normal distribution of our model.

The results of these tests are displayed in Figure \ref{postpred}.  The observed data's $f_{1\sigma}$ statistic (top) falls at the 63rd and 83rd percentiles (i.e.\ within ``one sigma") of the distributions for datasets generated from the individual posteriors and the population-wide posteriors, respectively.  Therefore, the model is fully consistent with the data for the aspect of the model that this statistic tests: the inferred intrinsic scatter of the M-R relation.  On the other hand, the observed data's $f_\mu$ statistic falls at the 2nd and 0.2th percentile of the distributions.  This test is sensitive to the shape of the distribution around the mean relation, and thereby probes the appropriateness of our assumption of a normal distribution in Eqn \ref{MR3}.  The fact that the data are marginally inconsistent with this aspect of the model reveals that this is an area of improvement for future work (see \S \ref{caveats} for a further discussion about this). 

The differences in the model-data fit implied by these two statistics illustrates how careful one must be in performing model checking and interpreting the results.  Instead of asking ``is the model consistent with the data?", a more well-posed question would be ``in what ways are the model consistent with the data?", especially as these statistical models become more complex to accommodate more sources of uncertainty and more realistic physics.  For the model at hand, we address this question by returning to our purpose.  We wanted to explore the need for intrinsic scatter in the M-R relation; therefore, we care most about the quality of the fit with respect to the spread of the M-R relation, i.e.\ the data-model consistency as quantified by $f_{1\sigma}$.  Because this fit is good, we are satisfied that our main result holds up to this further scrutiny.  The $f_\mu$ data-model discrepancy has implications for our choice to use a normal distribution for the probabilistic M-R relation, which we discuss in \S \ref{caveats}.

To wrap up our discussion on hierarchical Bayesian model checking, we note that the reader may find it surprising that the distribution of $f_{1\sigma}$ statistics peaks around $\sim 0.4$, given that the statistic was defined by the ``one-sigma" central region of the M-R relation.  The reason why the statistic does not instead cluster around 0.68 is due to a well-known feature of hierarchical modeling called shrinkage.  First worked out by \citet{Ste55} and developed within a Bayesian framework by \citet{Good65}, shrinkage refers to the tendency for individual parameter values at intermediate levels of hierarchical models (such as $M_t^{(i)},R_t^{(i)}$) to cluster more closely together, or ``shrink" toward their mean, than if the parameters had been analyzed completely independently of each other (we see this shrinking visually by comparing the gray and red lines in the right panel of Figure \ref{plotMRs}).  This occurs because, by design, the hierarchical structure of the model causes these individual parameters to be related to each other, which then enables information about one parameter to influence our inference for another parameter.  This additional information, provided solely by the hierarchical structure, causes the variance among the population to decrease relative to the case where the structure, and thus the information, was not available.  This decrease in variance is perceived as a ``shrinking" towards the mean.

We can understand how shrinkage manifests for the M-R relation by returning to the forward model defined in Eqn \ref{statmod}.  First we note that it is the true masses and radii, not the observed masses and radii, which are drawn from the M-R relation.  Therefore, it is the $M_t^{(i)}$ and $R_t^{(i)}$ values which would produce $f_{1\sigma}=0.68$.  However, we don't actually observe $M_t^{(i)}$ and $R_t^{(i)}$ outright; we observe them convolved with some error $\sigma_{ob}^{(i)}$.  Because the convolution of a normal distribution with another normal produces a wider normal distribution, we would indeed expect more of the observed mass and radius points to fall outside of the ``one-sigma" bounds defined by the top level of the model.  With this additional insight into the nature of hierarchical models, we understand that this behavior is not only consistent with what we see, but expected.

\subsection{Caveats and Future Work} \label{caveats}

As discussed in \S \ref{models}-\ref{data}, we made a number of assumptions and modeling choices to facilitate a straightforward investigation into the need for an intrinsic scatter term in the sub-Neptune M-R relation.  Some of these choices, such as parameterizing the relation with a power law or using a normal distribution to characterize the scatter, were driven by convenience and familiarity rather than by physics.  In particular, we chose to use a power law in order facilitate direct comparisons between our results and those in the literature, and we chose the normal distribution because it directly parameterizes the intrinsic scatter in the population rather than relying on a transformation from a more obscure distribution to derive the population variance.  In general, our philosophy was to limit the number of free parameters in our model as much as possible in order to maximize the information content per parameter.  

That said, these parameterizations are by no means the only ones that could be reasonably used.  These approximations can and should be revisited in future work, especially as more data becomes available and we begin to describe more subtle features in the M-R relation.  An important aspect of these studies will be model-checking to inform choices for parameterizing the M-R relation (see \S \ref{modelcheck}) and performing quantitative model comparisons.  For now, we emphasize that the inclusion of any kind of probability distribution to account for intrinsic scatter represents a significant improvement over prior work, and we leave testing different distributions for future studies.

A related parameterization issue is our choice not to include any other planet properties into this M-R relation, such as a dependence on orbital period.  It is entirely plausible on both theoretical and observational grounds that the M-R relation at short periods may be different from that at long periods.  Theoretically, photoevaporation is likely to have eroded planets on short orbits (see, for example, \citealt{Mur09,Lop12,Howe15}), thereby causing the population of highly irradiated planets to be denser on average; alternatively, migration could have produced a mass-dependent stopping location given a certain structure for the inner regions of the disk (e.g. \citealt{Ben11}) or tidal circularization could have produced a density-period correlation for the shortest orbits \citep{Barn14}.  Observationally, we see a suggestive dearth of $> 3$ \Rearth\ \Kep planet candidates and lower-mass RV planets at short orbital periods \citep{Bea13} and a period dependence in the marginalized, bias-corrected \Kep radius distribution \citep{You11,How12,Mor14}.  These observations hint at the potential for interesting features in the joint density-period distribution.  

While these results provide strong motivation for further investigations into period-dependent M-R relations, there will be numerous details to address in fitting the more complex period-dependent statistical model to the data.  We choose to leave these analyses for future work given the number of tests that we have already performed here (choice of parameterization for the intrinsic scatter, RV vs.\ TTV masses, and different radius ranges).  In particular, this future work will need to robustly account for the systematic biases between the different mass measurement methods to make sure that our inferences are not driven by arbitrary choices for the data we use (or if these choices prove to be unavoidable, to quantify their effect).  In general, future studies will need to perform model comparison to identify the most useful parameterizations given the data.  These two efforts pose cutting-edge statistical challenges that are worthy of separate investigations in and of themselves, and so we reserve them for follow-up studies that build on the results presented here.

Returning to our current statistical model for the M-R relation, our third major modeling choice after the power law and Gaussian astrophysical scatter was using Gaussian errors for the observed mass and radius measurements, as denoted in the last line of our statistical model (Eqn \ref{statmod}).  The most accurate hierarchical analyses include the actual likelihood used to infer these planetary parameters from the lower-level photometric and spectroscopic data, rather than assume such a heavily simplified functional form as we did.  Incorporating observers' true likelihoods is important to capture important correlations that exist in the measurement analysis and to use all of the information contained in the lower-level data.  Ideally, future planet discovery papers will make their full, joint likelihood distributions available in addition to the reported ``best-fit" value; depending on the type of analysis that observers use to make measurements from their data, this means providing either log-likelihood (e.g. $\chi^2$) contours over all of the parameters that were considered or providing posterior samples along with detailed information about the choice of priors.  Unfortunately, this information is not publicly available, and so we cannot use it.  Therefore, we follow the convention established by WM14 in using a normal distribution to represent the marginalized mass or radius likelihood.

Another notable approximation that we have made arises from the difference between the graphical model describing our M-R relation (Figure \ref{MRstruct}) and the way in which we have implemented it (Eqn \ref{statmod}).  The difference between the two is subtle: by conditioning on $\sigma_{Mob}^{(i)}$ in the lowest level of the model, we are assuming that the observed mass measurement uncertainty is independent from the measurement itself.  In reality, the calculation of both the uncertainty and the measured value are produced by the same modeling process and thus are correlated in a potentially nontrivial way\footnote{At first glance this may seem like an inconsequential difference considering the other modeling assumptions we have made, but the assumption of independence becomes problematic when the processes of detection and measurement use the same data and when the population analyses use regions of parameter space where the detection efficiency starts to drop (see \citet{Lor95} for the technical details).   We do not correct for detection bias in this paper, so are unaffected by this problem.}.   In practice, we had no other choice than to assume independence because the nature of this correlation is rarely, if ever, published.    









An important caveat about our results is that we ignore selection effects, primarily due to the inability to model the human decisions which affected the ground-based follow-up observing campaigns.  As discussed in \S \ref{data}, we would ideally have a uniform sample of masses and radii which were analyzed in the same way, and have well-characterized the selection effects and detection efficiencies.  Unfortunately, this is simply not possible at the present moment.  The sample is highly heterogeneous, a compilation of many observing teams'  programmatic, yearly, and nightly priorities that are not communicated in the literature and therefore cannot be accurately modeled.  There is significant, difficult work still to be done to understand the extent to which the follow-up process shapes the observed mass-radius space and therefore influence our inferences for the underlying M-R relation.

This is particularly important for interpreting the apparent discrepancy between the population of TTV-measured masses and the population of RV-measured masses.  In \S \ref{diffdat} we corroborate the systematic density difference between TTV and RV planets that had been noticed by \citet{Jon14} and \citet{Wei14}, as we find a $\sim$ two ``sigma" offset in the two datasets' M-R relation parameters.  Since we do not attempt to model the selection effects that are involved, we cannot distinguish how much of this offset is created by observational bias and how much is due to an intrinsic difference in the density distribution.  If the latter turns out to be the driving factor behind this apparent discrepancy, disentangling the underlying reason for this astrophysical density difference from the current list of features which distinguish the two samples will have numerous implications for planet formation and evolution.

\section{Conclusions} \label{Conclu}

In this paper we have defined and constrained a probabilistic mass-radius relationship for sub-Neptune planets (Eqn \ref{MR3} with parameter values in Table \ref{MRres} and the density constraint provided in Eqn \ref{maxM}).  In particular, we demonstrate that there is intrinsic, astrophysical scatter in this relation, and that, except for the smallest planets, this scatter is nonzero for all considered datasets.  For the first time in the exoplanet literature, we display the uncertainties in the M-R relation parameters through posterior distributions and explain how to properly incorporate these uncertainties into a predictive distribution of masses for individual planets.  This M-R relation will be useful for anyone who wishes to perform large-scale dynamical or planet formation studies with the \Kep planet candidates. 

More broadly, this work provides a framework for further analyses of the M-R relation and its probable dependencies on period and stellar properties.  Here we have demonstrated how to develop and apply a statistical model that incorporates measurement uncertainties into population-wide inference of the sub-Neptune M-R relation and that directly produces estimates of the uncertainty of the inferred parameters.  This method is advantageous because it is quantitative and easily generalizable to include additional variables which may be important in the underlying M-R relation that we are trying to model, such as incident flux (\S \ref{caveats}) or different stellar masses or metallicities.  We do not investigate these possibilities here, as we wish to begin this broader effort with the simplest reasonable statistical model.   Nevertheless, searching for additional dependencies in the M-R relation is an important endeavor in order to understand which physical processes shape the super-Earth population.  With this work we establish both a point of comparison and a framework for these further studies.

\acknowledgments

We thank Tom Loredo for lending his statistical expertise to the first version of this paper, which has facilitated increased clarity on some of the technical details of our modeling.  We also thank the anonymous referee for suggesting additional points of discussion to provide crucial context to this work.  This material was based upon work partially supported by the National Science Foundation under Grant DMS-1127914 to the Statistical and Applied Mathematical Sciences Institute, which organized a 3-week workshop on the Statistical Analysis of Kepler data during which the authors developed a preliminary hierarchical Bayesian model for this problem.  This research was also supported in part by the NSF Grant No. PHY11-25915 to the UCSB Kavli Institute for Theoretical Physics, which brought together the authors for final work on this project during its Dynamics and Evolution of Earth-like Planets program.  Any opinions, findings, and conclusions or recommendations expressed in this material are those of the author(s) and do not necessarily reflect the views of the National Science Foundation.

AW's financial support during this investigation was primarily provided by the National Science Foundation Graduate Research Fellowship under Grant No. 0809125, the UCSC Graduate Division's Eugene Cota-Robles Fellowship, with support during the referee process by the National Science Foundation under Award No. 1501440.  LAR gratefully acknowledges support provided by NASA through Hubble Fellowship grant \#HF-51313 awarded by the Space Telescope Science Institute, which is operated by the Association of Universities for Research in Astronomy, Inc., for NASA, under contract NAS 5-26555.  EBF was supported in part by NASA \textit{Kepler} Participating Scientist Program awards NNX12AF73G \& NNX14AN76G, NASA Origins of Solar Systems awards NNX13AF61G \& NNX14AI76G, and NASA Exoplanet Research Program award NNX15AE21G.  The Center for Exoplanets and Habitable Worlds is supported by the Pennsylvania State University, the Eberly College of Science, and the Pennsylvania Space Grant Consortium.  This research has made use of the NASA Exoplanet Archive, which is operated by the California Institute of Technology, under contract with the National Aeronautics and Space Administration under the Exoplanet Exploration Program.  This paper includes data collected by the Kepler mission. Funding for the Kepler mission is provided by the NASA Science Mission directorate.

{\it Facilities:} \facility{Kepler}.

\clearpage
\LongTables
\begin{deluxetable*}{lrrrllccc}
\tabletypesize{\footnotesize}
\tablecolumns{9}
\tablewidth{0pt}
\tablecaption{Masses and Radii of Small Planets \label{dataset}}
\tablehead{
\colhead{Planet Name} & \colhead{Period} & \colhead{$M_{obs}$}  & \colhead{$\sigma_{Mobs}$}  & \colhead{$R_{obs}$}  & \colhead{$\sigma_{Robs}$} &  \colhead{First} & \colhead{Mass, Radius} & \colhead{Note} \\
 & \colhead{(days)} & \colhead{(\Mearth)} & \colhead{(\Mearth)} & \colhead{(\Rearth)} & \colhead{(\Rearth)} &  \colhead{Reference} & \colhead{Reference} & 
}
\startdata
55 Cnc e  &  0.737  &  8.09  &  0.26  &  2.17  &  0.098  &  McArthur(2004)  &  Nelson(2014), Gillon(2012)  &  0  \\
CoRoT-7 b  &  0.854  &  4.73  &  0.95  &  1.58  &  0.064  &  Queloz(2009); Leger(2009)  &  Barros(2014)  &  0  \\
GJ 1214 b  &  1.580  &  6.45  &  0.91  &  2.65  &  0.09  &  Charbonneau(2009)  &  Carter(2011)  &  0,1  \\
GJ 3470 b  &  3.337  &  13.73  &  1.61  &  3.88  &  0.32  &  Bonfils(2012)  &  Biddle(2014)  &  0  \\
HD 97658 b  &  9.491  &  7.87  &  0.73  &  2.34  &  0.16  &  Howard(2011)  &  Dragomir(2013)  &  0,1  \\
HIP 116454 b  &  9.12\phantom{0}  &  11.82  &  1.33  &  2.53  &  0.18  &  Vanderburg(2015)  &  Vanderburg(2015)  &  0  \\
Kepler-10 b  &  0.837  &  3.33  &  0.49  &  1.47  &  0.02  &  Batalha(2011)  &  Dumusque(2014)  &  0  \\
Kepler-10 c  &  45.294  &  17.2\phantom{0}  &  1.9\phantom{0}  &  2.35  &  0.06  &  Batalha(2011)  &  Dumusque(2014)  &  0  \\
Kepler-19 b  &  9.287  &  ---\phantom{0}  &  20.3\phantom{0}  &  2.21  &  0.048  &  Borucki(2011)  &  Ballard(2011)  &  0,4  \\
Kepler-20 b  &  3.696  &  8.7\phantom{0}  &  2.2\phantom{0}  &  1.91  &  0.16  &  Borucki(2011)  &  Gautier(2012)  &  0  \\
Kepler-20 c  &  10.854  &  16.1\phantom{0}  &  3.5 \phantom{0} &  3.07  &  0.25  &  Borucki(2011)  &  Gautier(2012)  &  0  \\
Kepler-20 d  &  77.612  &  ---\phantom{0}  &  20.1\phantom{0}  &  2.75  &  0.23  &  Borucki(2011)  &  Gautier(2012)  &  0,4  \\
Kepler-20 e  &  6.098  &  ---\phantom{0}  &  3.08  &  0.868  &  0.08  &  Borucki(2011)  &  Fressin(2012)  &  0,4  \\
Kepler-20 f  &  19.58\phantom{0}  &  ---\phantom{0}  &  14.3\phantom{0}  &  1.03  &  0.11  &  Borucki(2011)  &  Fressin(2012)  &  0,4  \\
Kepler-21 b  &  2.786  &  ---\phantom{0}  &  10.4\phantom{0}  &  1.635  &  0.04  &  Borucki(2011)  &  Howell(2012)  &  0,4  \\
Kepler-25 b  &  6.239  &  9.60  &  4.20  &  2.71  &  0.05  &  Borucki(2011)  &  Marcy(2014)  &  0,1  \\
Kepler-37 b  &  13.367  &  2.78  &  3.70  &  0.32  &  0.02  &  Borucki(2011)  &  Marcy(2014)  &  0,1  \\
Kepler-37 c  &  21.302  &  3.35  &  4.00  &  0.75  &  0.03  &  Borucki(2011)  &  Marcy(2014)  &  0,1  \\
Kepler-37 d  &  39.792  &  1.87  &  9.08  &  1.94  &  0.06  &  Borucki(2011)  &  Marcy(2014)  &  0,1  \\
Kepler-48 b  &  4.778  &  3.94  &  2.10  &  1.88  &  0.10  &  Borucki(2011)  &  Marcy(2014)  &  0,1  \\
Kepler-48 c  &  9.674  &  14.61  &  2.30  &  2.71  &  0.14  &  Borucki(2011)  &  Marcy(2014)  &  0,1  \\
Kepler-48 d  &  42.896  &  7.93  &  4.60  &  2.04  &  0.11  &  Borucki(2011)  &  Marcy(2014)  &  0,1  \\
Kepler-62 b  &  5.715  &  ---\phantom{0}  &  9\phantom{.00}  &  1.31  &  0.04  &  Borucki(2011)  &  Borucki(2013)  &  0,4  \\
Kepler-62 c  &  12.44\phantom{0}  &  ---\phantom{0}  &  4\phantom{.00}  &  0.54  &  0.03  &  Borucki(2013)  &  Borucki(2013)  &  0,4  \\
Kepler-62 d  &  18.164  &  ---\phantom{0}  &  14\phantom{.00}  &  1.95  &  0.07  &  Borucki(2011)  &  Borucki(2013)  &  0,4  \\
Kepler-62 e  &  122.39\phantom{0}  &  ---\phantom{0}  &  36\phantom{.00}  &  1.61  &  0.05  &  Borucki(2011)  &  Borucki(2013)  &  0,4  \\
Kepler-62 f  &  267.29\phantom{0}  &  ---\phantom{0}  &  35\phantom{.00}  &  1.41  &  0.07  &  Borucki(2013)  &  Borucki(2013)  &  0,6  \\
Kepler-68 b  &  5.399  &  5.97  &  1.70  &  2.33  &  0.02  &  Borucki(2011)  &  Marcy(2014)  &  0,3  \\
Kepler-68 c  &  9.605  &  2.18  &  3.50  &  1.00  &  0.02  &  Batalha(2013)  &  Marcy(2014)  &  0,3  \\
Kepler-78 b  &  0.354  &  1.69  &  0.41  &  1.20  &  0.09  &  Sanchis-Ojeda(2013a)  &  Howard(2013)  &  0,1  \\
Kepler-89 b  &  3.743  &  10.50  &  4.60  &  1.71  &  0.16  &  Borucki(2011)  &  Weiss(2013)  &  0,1  \\
Kepler-93 b  &  4.727  &  4.02  &  0.68  &  1.48  &  0.019  &  Borucki(2011)  &  Dressing(2015)  &  0  \\
Kepler-94 b  &  2.508  &  10.84  &  1.40  &  3.51  &  0.15  &  Borucki(2011)  &  Marcy(2014)  &  0,1  \\
Kepler-95 b  &  11.523  &  13.00  &  2.90  &  3.42  &  0.09  &  Borucki(2011)  &  Marcy(2014)  &  0,1  \\
Kepler-96 b  &  16.238  &  8.46  &  3.40  &  2.67  &  0.22  &  Borucki(2011)  &  Marcy(2014)  &  0,1  \\
Kepler-97 b  &  2.587  &  3.51  &  1.90  &  1.48  &  0.13  &  Borucki(2011)  &  Marcy(2014)  &  0,1  \\
Kepler-98 b  &  1.542  &  3.55  &  1.60  &  1.99  &  0.22  &  Borucki(2011)  &  Marcy(2014)  &  0,1  \\
Kepler-99 b  &  4.604  &  6.15  &  1.30  &  1.48  &  0.08  &  Borucki(2011)  &  Marcy(2014)  &  0,1  \\
Kepler-100 b  &  6.887  &  7.34  &  3.20  &  1.32  &  0.04  &  Borucki(2011)  &  Marcy(2014)  &  0,1  \\
Kepler-100 c  &  12.816  &  0.85  &  4.00  &  2.20  &  0.05  &  Borucki(2011)  &  Marcy(2014)  &  0,1  \\
Kepler-100 d  &  35.333  &  -4.36  &  4.10  &  1.61  &  0.05  &  Borucki(2011)  &  Marcy(2014)  &  0,1  \\
Kepler-101 c  &  6.03\phantom{0}  &  ---\phantom{0}  &  9\phantom{.00}  &  1.25  &  0.18  &  Borucki(2011)  &  Bonomo(2014)  &  0,5  \\
Kepler-102 d  &  10.312  &  3.80  &  1.80  &  1.18  &  0.04  &  Borucki(2011)  &  Marcy(2014)  &  0,1  \\
Kepler-102 e  &  16.146  &  8.93  &  2.00  &  2.22  &  0.07  &  Borucki(2011)  &  Marcy(2014)  &  0,1  \\
Kepler-102 f  &  27.454  &  0.62  &  3.30  &  0.88  &  0.03  &  Borucki(2011)  &  Marcy(2014)  &  0,1  \\
Kepler-102 b  &  5.287  &  0.41  &  1.60  &  0.47  &  0.02  &  Borucki(2011)  &  Marcy(2014)  &  0,1  \\
Kepler-102 c  &  7.071  &  -1.58  &  2.00  &  0.58  &  0.02  &  Borucki(2011)  &  Marcy(2014)  &  0,1  \\
Kepler-103 b  &  15.965  &  14.11  &  4.70  &  3.37  &  0.09  &  Borucki(2011)  &  Marcy(2014)  &  0,1  \\
Kepler-106 b  &  6.165  &  0.15  &  2.80  &  0.82  &  0.11  &  Borucki(2011)  &  Marcy(2014)  &  0,1  \\
Kepler-106 c  &  13.571  &  10.44  &  3.20  &  2.50  &  0.32  &  Borucki(2011)  &  Marcy(2014)  &  0,1  \\
Kepler-106 d  &  23.980  &  -6.39  &  7.00  &  0.95  &  0.13  &  Batalha(2013)  &  Marcy(2014)  &  0,1  \\
Kepler-106 e  &  43.844  &  11.17  &  5.80  &  2.56  &  0.33  &  Borucki(2011)  &  Marcy(2014)  &  0,1  \\
Kepler-109 b  &  6.482  &  1.30  &  5.40  &  2.37  &  0.07  &  Borucki(2011)  &  Marcy(2014)  &  0,1  \\
Kepler-109 c  &  21.223  &  2.22  &  7.80  &  2.52  &  0.07  &  Borucki(2011)  &  Marcy(2014)  &  0,1  \\
Kepler-113 b  &  4.754  &  7.10  &  3.30  &  1.82  &  0.05  &  Borucki(2011)  &  Marcy(2014)  &  0,1  \\
Kepler-113 c  &  8.925  &  -4.60  &  6.20  &  2.19  &  0.06  &  Borucki(2011)  &  Marcy(2014)  &  0,1  \\
Kepler-131 b  &  16.092  &  16.13  &  3.50  &  2.41  &  0.20  &  Borucki(2011)  &  Marcy(2014)  &  0,1  \\
Kepler-131 c  &  25.517  &  8.25  &  5.90  &  0.84  &  0.07  &  Batalha(2013)  &  Marcy(2014)  &  0,1  \\
Kepler-406 b  &  2.426  &  4.71  &  1.70  &  1.43  &  0.03  &  Borucki(2011)  &  Weiss(2014)  &  0,1  \\
Kepler-406 c  &  4.623  &  1.53  &  2.30  &  0.85  &  0.03  &  Batalha(2013)  &  Weiss(2014)  &  0,1  \\
Kepler-407 b  &  0.669  &  0.06  &  1.20  &  1.07  &  0.02  &  Borucki(2011)  &  Marcy(2014)  &  0,1  \\
Kepler-409 b  &  68.958  &  2.69  &  6.20  &  1.19  &  0.03  &  Batalha(2013)  &  Marcy(2014)  &  0,1  \\
Kepler-4 b  &  3.213  &  24.47  &  3.81  &  4.00  &  0.21  &  Borucki(2010)  &  Borucki(2010)  &    \\
GJ 436 b  &  2.64\phantom{0}  &  25.4\phantom{0}  &  2.1\phantom{0}  &  4.10  &  0.16  &  Butler(2004)  &  Lanotte(2014)  &    \\
Kepler-89 c  &  10.42\phantom{0}  &  15.6\phantom{0}  &  10.6\phantom{0}  &  4.32  &  0.41  &  Batalha(2013)  &  Weiss(2013)  &    \\
HAT-P-11 b  &  4.888  &  25.74  &  2.86  &  4.73  &  0.157  &  Bakos(2010)  &  Bakos(2010)  &    \\
CoRoT-22 b  &  9.756  &  ---\phantom{0}  &  35\phantom{.00}  &  4.88  &  0.28  &  Moutou(2014)  &  Moutou(2014)  &  4  \\
Kepler-103 c  &  179.61\phantom{0}  &  36.1\phantom{0}  &  25.2\phantom{0}  &  5.14  &  0.14  &  Borucki(2011)  &  Marcy(2014)  &    \\
Kepler-101 b  &  3.488  &  51.1\phantom{0}  &  4.9\phantom{0}  &  5.77  &  0.82  &  Borucki(2011)  &  Bonomo(2014)  &    \\
Kepler-63 b  &  9.43\phantom{0}  &  ---\phantom{0}  &  95\phantom{.00}  &  6.1  &  0.2  &  Borucki(2011)  &  Sanchis-Ojeda(2013b)  &  6  \\
HAT-P-26 b  &  4.235  &  18.75  &  2.23  &  6.33  &  0.58  &  Hartman(2011)  &  Hartman(2011)  &    \\
CoRoT-8 b  &  6.212  &  69.92  &  9.53  &  6.39  &  0.22  &  Borde(2010)  &  Borde(2010)  &    \\
Kepler-89 e  &  54.32\phantom{0}  &  35\phantom{.00}  &  23\phantom{.00}  &  6.56  &  0.62  &  Batalha(2013)  &  Weiss(2013)  &    \\
Kepler-11 b  &  10.304  &  1.90  &  1.2\phantom{0}  &  1.80  &  0.04  &  Lissauer(2011)  &  Lissauer(2013)  &  1,2  \\
Kepler-11 c  &  13.024  &  2.90  &  2.3\phantom{0}  &  2.87  &  0.06  &  Lissauer(2011)  &  Lissauer(2013)  &  1,2  \\
Kepler-11 d  &  22.684  &  7.30  &  1.2\phantom{0}  &  3.12  &  0.07  &  Lissauer(2011)  &  Lissauer(2013)  &  1,2  \\
Kepler-11 f  &  46.689  &  2.00  &  0.9\phantom{0}  &  2.49  &  0.06  &  Lissauer(2011)  &  Lissauer(2013)  &  1,2  \\
Kepler-11 g  &  118.38\phantom{0}  &  ---\phantom{0}  &  25\phantom{.00}  &  3.33  &  0.07  &  Lissauer(2011)  &  Lissauer(2013)  &  2,4  \\
Kepler-18 b  &  3.505  &  6.9\phantom{0}  &  3.4\phantom{0}  &  2.00  &  0.100  &  Borucki(2011)  &  Cochran(2011)  &  1,2  \\
Kepler-30 b  &  29.334  &  11.3\phantom{0}  &  1.4\phantom{0}  &  3.90  &  0.200  &  Borucki(2011)  &  Sanchis-Ojeda(2012)  &  1,2  \\
Kepler-36 b  &  13.840  &  4.45  &  0.30  &  1.486  &  0.035  &  Carter(2012)  &  Carter(2012)  &  1,2  \\
Kepler-36 c  &  16.239  &  8.08  &  0.53  &  3.679  &  0.054  &  Borucki(2011)  &  Carter(2012)  &  1,2  \\
Kepler-79 b  &  13.485  &  10.9\phantom{0}  &  6.7\phantom{0}  &  3.47  &  0.07  &  Borucki(2011)  &  Jontof-Hutter(2014)  &  1,2  \\
Kepler-79 c  &  27.403  &  5.9\phantom{0}  &  2.1\phantom{0}  &  3.72  &  0.08  &  Borucki(2011)  &  Jontof-Hutter(2014)  &  1,2  \\
Kepler-79 e  &  81.066  &  4.1\phantom{0}  &  1.2\phantom{0}  &  3.49  &  0.14  &  Batalha(2013)  &  Jontof-Hutter(2014)  &  1,2  \\
Kepler-88 b  &  10.954  &  8.7\phantom{0}  &  2.5\phantom{0} &  3.78  &  0.38  &  Borucki(2011)  &  Nesvorny(2013)  &  2  \\
Kepler-138 c  &  13.782  &  3.83  &  1.39  &  1.610  &  0.160  &  Borucki(2011)  &  Kipping(2014)  &  2  \\
Kepler-138 d  &  23.089  &  1.01  &  0.38  &  1.610  &  0.160  &  Borucki(2011)  &  Kipping(2014)  &  2  \\
Kepler-289 b  &  34.545  &  7.3\phantom{0}  &  6.8\phantom{0}  &  2.15  &  0.1  &  Borucki(2011)  &  Schmitt(2014)  &  2  \\
Kepler-289 d  &  66.063  &  4.0\phantom{0}  &  0.9\phantom{0}  &  2.68  &  0.17  &  Borucki(2011)  &  Schmitt(2014)  &  2
\enddata
\tablecomments{
\vspace{-3mm}
\begin{minipage}[t]{1.8\textwidth}
\begin{enumerate}
\setcounter{enumi}{-1}
\item Included in baseline dataset, which consists of RV masses (see \S \ref{data}). 
\item Mass, radius values and their error bars are unchanged (within rounding error) from WM14.
\item Mass measured by fitting the observed TTVs to N-body integrations of the system.
\item The Kepler-68 planets were repeated twice in the WM14 dataset, so we use the \citet{Mar14} values.
\item The $\sigma_{Mobs}$ column contains the $2\sigma$ upper limit as reported in the second reference.
\item Only a $1\sigma$ upper limit of 3.78 was given, and no posteriors were shown; in this analysis, we set the $2\sigma$ upper limit at 9 \Mearth\ to include 1.8 m/s uncertainty quoted in RV semi-amplitude for the larger Kepler-101 b.
\item The $2\sigma$ upper limit is interpolated from given $1\sigma$ and $3\sigma$ upper limits.
\end{enumerate}
\end{minipage}
}
\end{deluxetable*}
\clearpage

\end{document}